\documentclass[a4paper,oribibl]{llncs}
\usepackage{etex}
\usepackage[table]{xcolor} 
\usepackage[utf8]{inputenc}
\usepackage{amsmath,amssymb}
\usepackage{float}
\usepackage{xspace}
\usepackage{calc}
\usepackage{caption}
\usepackage{subcaption}
\usepackage{enumitem}
\usepackage{booktabs}
\usepackage{tabu}
\usepackage{multirow}
\usepackage{url}
\usepackage{xspace}
\usepackage{graphicx} 
\usepackage{longtable} 
\usepackage{pifont}
\usepackage{pgfplots}
\usepackage{pgfkeys}
\usepackage{tikz}
\usetikzlibrary{shapes,shapes.symbols,arrows,decorations,decorations.text,decorations.pathmorphing,
decorations.pathreplacing,decorations.shapes,backgrounds,fit,positioning,calc,arrows,automata}
\usetikzlibrary{calc, backgrounds}
\usepackage{listings} 
\usepackage{placeins}
\usepackage{gensymb} 
\usepackage[disable]{todonotes}
\usepackage{hyperref} 
\usepackage[toc,page]{appendix}
\usepackage{mathtools}
\usepackage[linesnumbered,lined,ruled]{algorithm2e}
\renewcommand{\gets}{\coloneqq}
\usepackage{stmaryrd}
\usepackage{blindtext}

\definecolor{orange}   		{RGB}{255,128,0}
\definecolor{darkred}  		{RGB}{128,0,0}
\definecolor{darkgreen}		{RGB}{0,128,0}
\definecolor{dirtgreen}		{RGB}{180,210,180}
\definecolor{mixgreen}		{RGB}{34,139,34}
\definecolor{darkblue} 		{RGB}{0,0,128}
\definecolor{darkpurple}	{RGB}{160,32,240}
\definecolor{lightpurple}	{RGB}{180,180,210}
\definecolor{bl1}   		{RGB}{204,204,255}
\definecolor{bl2}   		{RGB}{128,128,255}
\definecolor{bl3}   		{RGB}{140,160,255}

\definecolor{gr1}		{RGB}{250, 250, 250}
\definecolor{gr2}		{RGB}{229, 229, 229}
\definecolor{gr3}		{RGB}{212, 212, 212}
\definecolor{gr4}		{RGB}{204, 204, 204}

\definecolor{g1}		{RGB}{215,25,28} 
\definecolor{g2}		{RGB}{253,174,97} 
\definecolor{g3}		{RGB}{255,255,191} 
\definecolor{g4}		{RGB}{171,217,233} 
\definecolor{g5}		{RGB}{44,123,182} 

\definecolor{s1}		{RGB}{228,26,28}
\definecolor{s2}		{RGB}{55,126,184}
\definecolor{s3}		{RGB}{77,175,74}
\definecolor{s4}		{RGB}{152,78,163}
\definecolor{s5}		{RGB}{255,127,0}
\definecolor{s6}		{RGB}{255,255,51}
\definecolor{s7}		{RGB}{166,86,40}
\definecolor{s8}		{RGB}{247,129,191}
\definecolor{s9}		{RGB}{153,153,153}

\colorlet{codegreen}		{mixgreen}
\colorlet{codepurple}		{darkpurple}
\colorlet{codehighlight}	{gr2}

\colorlet{stmtcolor}		{gr2} 
\colorlet{prg}				{gr2} 
\colorlet{ltl}				{g4} 
\colorlet{notIC}			{bl3} 
\colorlet{stateass}			{g2}

\colorlet{outlineblue}		{g5}
\colorlet{fillblue}			{g4}
\colorlet{darkback}			{gr2}
\colorlet{lightback}		{gr1}

\colorlet{source}			{gr1}
\colorlet{controller}		{gr1}
\colorlet{thirdparty}		{gr1} 
\colorlet{core}				{gr1}
\colorlet{libs}				{gr1}
\colorlet{source}			{gr1}
\colorlet{analyzer}			{gr1}
\colorlet{generator}		{gr1}
\colorlet{output}			{gr1}

\hypersetup{colorlinks  
  ,pdfpagemode=UseNone
  ,linkcolor=black  
  ,filecolor=black 
  ,urlcolor=black
  ,citecolor=black
} 

\newlength{\imgwidth}

\newcommand{\prog}{\ensuremath{\mathcal{P}}\xspace}
\newcommand{\progn}[1]{\ensuremath{\mathcal{P}_{#1}}\xspace}
\newcommand{\pprog}{\ensuremath{\mathcal{P^{\#}}}\xspace}
\newcommand{\pprogtau}{\ensuremath{\mathcal{P_{\tau}^{\#}}}\xspace}

\newcommand{\Stmt}{\ensuremath{\mathit{Stmt}}\xspace}
\newcommand{\States}{\ensuremath{\mathit{S}}\xspace}

\newcommand{\var}{\ensuremath{\mathit{Var}}\xspace}
\newcommand{\ipfname}{\mathcal{I}}
\newcommand{\ipf}[2]{\ensuremath{\ipfname(\texttt{#1})(#2)}\xspace}

\newcommand{\atm}[1]{\ensuremath{\mathcal{A}_{#1}}\xspace}
\newcommand{\langatm}[1]{\ensuremath{\mathcal{L}(\atm{#1})}\xspace}

\newcommand{\hoare}[3]{\ensuremath{\{#1\} \; #2 \; \{#3\}}\xspace}

\tikzstyle{trans} = [->,>=stealth]
\tikzstyle{smallnode} = [inner sep=1.4]
\tikzstyle{st} = [%
	font=\ttfamily, %
	shape=rectangle, %
	rounded corners=.5em, %
	fill=prg, %
	inner xsep=.3em,%
	inner ysep=0em, %
	text height=2ex,%
	text depth=.6ex %
]

\tikzstyle{formula} = [%
	shape=rectangle, %
	line width=1pt, %
	draw=#1,%
	fill=#1!40, %
	inner xsep=.3em,%
	inner ysep=.0em,%
	text height=2ex,%
	text depth=.6ex %
]

\tikzstyle{check} = [%
	draw=black, %
	fill=white, %
	text centered, %
	minimum width=5cm,%
	minimum height=8mm,%
	outer sep=1mm,%
]

\tikzstyle{flow} = [%
	->, %
	very thick, %
	inner sep=0mm,%
] 

\tikzstyle{formula2} = [%
	shape=rectangle,%
	line width=1pt,%
	draw=#1,%
	fill=#1!40,%
	inner xsep=.3em,%
	inner ysep=.0em%
]

\newcommand{\stfootcol}[2]{\tikz[baseline]{\node[st,fill=#2] at (0,.64ex){\hspace{.3em}\texttt{\strut\footnotesize #1}\hspace{.3em}\strut};}}

\newcommand{\safootcol}[2]{\tikz[baseline]{\node[formula=#2] at (0,.64ex){\hspace{.3em}\texttt{\strut\footnotesize #1}\hspace{.3em}\strut};}}
\newcommand{\sadyn}[1]{\tikz[baseline]{\node[formula2=stateass] at (0,.64ex){\hspace{.3em}\texttt{\strut\footnotesize #1}\hspace{.3em}\strut};}}

\newcommand{\st}{{s\!t}}

\newcommand{\stprog}[1]{\ensuremath{\protect\stfootcol{#1}{prg}}\xspace}
\newcommand{\saprog}[1]{\ensuremath{\protect\safootcol{#1}{stateass}}\xspace}

\newcommand\definetool[2]{\newcommand{#1}{\texorpdfstring{{\scshape #2}}{#2}\xspace}}
\definetool{\ultimate}     {Ultimate}
\definetool{\ltlautomizer} {Ultimate LTLAutomizer}
\definetool{\automizer}    {Ultimate Automizer}
\definetool{\cpachecker}   {CPAchecker}
\definetool{\blast}        {Blast}
\definetool{\slam}         {SLAM}
\newcommand{\svcomp}{{SV-COMP}\xspace}

\newcommand{\setAuto}{Default}
\newcommand{\setComp}{OCT+CON}
\newcommand{\setCompT}{OCT+CON Enh.}
\newcommand{\setInt}{INT}
\newcommand{\setIntT}{INT Enh.}
\newcommand{\setOct}{OCT}
\newcommand{\setOctT}{OCT Enh.}

\newcommand{\feas}{\mathit{feas}}
\newcommand{\tffeas}{T_\feas} 

\newcommand{\headcolor}{}

\newcommand{\header}[1]{{#1}\headcolor}

\newcommand{\refF}{\ensuremath{\textsf{generalize}}\xspace}

\newcommand{\dd}[1]{\todo[color=green!40]{#1}\xspace}
\newcommand{\mg}[1]{\todo[color=blue!40]{#1}\xspace}
\newcommand{\todoin}[1]{\todo[inline]{#1}}
\newcommand{\cn}{\todo{Citation needed}\xspace}
\newcommand{\ddin}[2]{\todo[inline,color=green!40,caption={#1}]{#2}}
\newcommand{\ddins}[1]{\ddin{#1}{#1}}
\newcommand{\mgin}[2]{\todo[inline,color=blue!40,caption={#1}]{#2}}
\newcommand{\mgins}[1]{\mgin{#1}{#1}}
\newcommand{\reviewer}[1]{\todo[color=yellow!40]{#1}\xspace}
\newcommand{\reviewerin}[2]{\todo[inline,color=yellow!40,caption={#1}]{#2}}
\newcommand{\reviewerins}[1]{\reviewerin{#1}{#1}}

\newcommand{\funcdef}[3]{\ensuremath{#1} : {#2} \to {#3}}
\newcommand{\loc}{\ensuremath{\ell}}
\newcommand{\locs}{\ensuremath{\mathit{Loc}}}
\newcommand{\trans}{\ensuremath{\delta}}
\newcommand{\dom}{\ensuremath{\mathcal{D}}}
\newcommand{\progdef}{\ensuremath{\prog = (\locs, \trans, \loc_{0})}\xspace}
\newcommand{\edgedef}{\ensuremath{\trans \subseteq \locs \times \Stmt \times \locs}}

\newcommand{\absstates}{\ensuremath{A}}

\newcommand{\widenop}{\ensuremath{\nabla}}

\makeatletter
\DeclareRobustCommand{\bigwidenop}{%
	\mathop{\vphantom{\sum}\mathpalette\bigwidenop@\relax}\slimits@
}

\newcommand{\bigwidenop@}[2]{%
	\vcenter{%
		\sbox\z@{$#1\sum$}%
		\hbox{\resizebox{.9\dimexpr\ht\z@+\dp\z@}{!}{$\m@th\nabla$}}%
	}%
}
\makeatother

\newcommand{\abspost}[2]{\ensuremath{{#1} \llbracket {#2} \rrbracket}}
\newcommand{\abspostdefault}{\ensuremath{\abspost{\cdot}{\cdot}}}
\newcommand{\abspostdef}[1][\lattice]{\ensuremath{\funcdef{\abspostdefault}{{#1}
\times \Stmt}{{#1}}}}

\newcommand{\absdomain}[1][A]{\ensuremath{{#1}^{\#}}}
\newcommand{\absdomaindef}[1][A]{\ensuremath{\absdomain[#1] := ({#1}, \widenop,
\abspostdefault)}}

\newenvironment{customlegend}[1][]{%
    \begingroup
    \csname pgfplots@init@cleared@structures\endcsname
    \pgfplotsset{#1}%
}{%
    \csname pgfplots@createlegend\endcsname
    \endgroup
}%

\def\addlegendimage{\csname pgfplots@addlegendimage\endcsname}

\pgfplotsset{every axis/.append style={thick}}

\definecolor{s1}{RGB}{228,26,28}\definecolor{s2}{RGB}{55,126,184}\definecolor{s3}{RGB}{77,175,74}\definecolor{s4}{RGB}{152,78,163}\definecolor{s5}{RGB}{255,127,0}\definecolor{s6}{RGB}{255,255,51}\definecolor{s7}{RGB}{166,86,40}\definecolor{s8}{RGB}{247,129,191}\definecolor{s9}{RGB}{153,153,153}\pgfplotsset{
    mark repeat/.style={
        scatter,
        scatter src=x,
        scatter/@pre marker code/.code={
            \pgfmathtruncatemacro\usemark{
                or(mod(\coordindex,#1)==0, (\coordindex==(\numcoords-1))
            }
            \ifnum\usemark=0
                \pgfplotsset{mark=none}
            \fi
        },
        scatter/@post marker code/.code={}
    }
}
\pgfplotsset{cycle list={%
{draw=s1,solid},
{draw=s2,dotted},
{draw=s3,dashed},
{mark repeat={10},draw=s4,solid,mark=star},
{mark repeat={10},draw=s5,solid,mark=triangle},
{mark repeat={10},draw=s6,solid,mark=diamond},
{mark repeat={10},draw=s7,solid,mark=x},
{mark repeat={10},draw=s8,solid,mark=|},
{mark repeat={10},draw=s9,solid,mark=10-pointed-star},
{mark repeat={10},draw=black,solid,mark=pentagon},
{mark repeat={10},draw=OliveGreen,solid,mark=o},
}}

\begin{document}

\title{\textsc{Refining Trace Abstraction using Abstract Interpretation}} 

\author{%
Marius Greitschus\and
Daniel Dietsch\and
Andreas Podelski
}
 
\institute{University of Freiburg, Germany}

\maketitle

\mg{16 pages, excluding bibliography, appendices}
\mg{Some of the minor comments by TACAS reviewer 3 already incorporated.}
\reviewer{$2^{\Stmt}$ should be $\Stmt^{*}$}

\begin{abstract}
The CEGAR loop in software model checking notoriously diverges when the
abstraction refinement procedure does not derive a loop invariant.
An abstraction refinement procedure based on an SMT solver is applied to a
trace, i.e., a restricted form of a program (without loops).
In this paper, we present a new abstraction refinement procedure that aims at
circumventing this restriction whenever possible.
We apply abstract interpretation to a program that we derive from the given
trace.
If the program contains a loop, we are guaranteed to obtain a loop invariant.
We call an SMT solver only in the case where the abstract interpretation returns
an indefinite answer.
That is, the idea is to use abstract interpretation and an SMT solver in tandem.
An experimental evaluation in the setting of trace abstraction indicates the
practical potential of this idea.
\end{abstract}

\section{Introduction}\label{sec:intro}
\todo{Say that our algorithm is implemented in \automizer but could be used for
any kind of CEGAR-based SW model checker.}
When trying to prove the correctness of a program, finding useful abstractions
in form of state assertions is the most important part of the
process~\cite{floyd1967assigning, hoare1969axiomatic}.
In this context, usefulness is about being able to prove correctness as
efficiently as possible.
Hence, in order to be able to analyze large programs, it is important to
find state assertions automatically.
The conflict between these two goals\mg{I don't see any conflict here. I don't
even see two goals. I see just one: Efficiently find SAs, i.e. automatically
for large programs.}\dd{automatically vs ``useful'': Humans can find small and
useful ones, but do not scale. Tools find many automatically, but they may not
be useful, i.e., divergence} gives rise to different techniqes for synthesizing
state assertions.
For example, abstract interpretation~\cite{DBLP:conf/popl/CousotC77} is a
well-known method for finding state assertions.
Abstract interpretation computes an over-approximation of a program's states by
using an up-front and largely program-independent abstraction. 
Many such abstractions exist (\cite{DBLP:conf/popl/CousotH78,
DBLP:conf/popl/SagivRW99, DBLP:conf/vmcai/SankaranarayananSM05,
DBLP:conf/vmcai/BeyerHMR07}) and all of them are useful, because they give rise
to different kinds of state assertions that can be used to prove the correctness
of different kinds of programs.
It is the strength of abstract interpretation that it always terminates and
always computes a fixpoint in the selected abstraction. 
If the program contains loops, the fixpoint computed for the loop head is also,
by definition, a loop invariant which allows for an easy abstraction of loops. 
While abstract interpretation scales favorably with the size of the
program, the computed over-approximation is often not precise enough to be
useful to prove the correctness of a program.

Another example to address this task is to use
software model checking tools like \blast~\cite{DBLP:journals/sttt/BeyerHJM07},
\slam~\cite{DBLP:conf/cav/BallR01}, and more recently,
\cpachecker~\cite{beyer2011cpachecker} and
\automizer~\cite{DBLP:conf/tacas/HeizmannCDEHLNSP13}, that follow the
counterexample-guided abstraction refinement (CEGAR)
approach~\cite{clarke2000counterexample}.
In CEGAR, an abstraction is continuously refined by synthesizing state
assertions from paths through the control flow graph of the program that 1) are
not contained in the current abstraction, 2) can reach an error location, and
3) are not executable.
By extracting state assertions from those paths, the abstraction can be refined
to fit the program at hand, which allows the user a greater amount of
flexibility in choosing her programs.\mg{I don't understand what the user has
to do with the process here. It was all automatic. And suddenly the user comes
into play?}\dd{The user has to select the abstraction in Ai because the
abstraction is program-independent}
Because the path analysis has to be precise, i.e., it has to ensure that paths
that represent real errors can be identified, it often produces state assertions
that are too strong to be loop invariants, in turn forcing the CEGAR algorithm
to unroll loops of the program.
If this happens, the algorithm may not be able to refine the abstraction at
all, e.g., because the loop of the analyzed program can be unrolled infinitely
often.

In this paper we propose a unification of both techniques, abstract
interpretation and CEGAR-based software model checking, such that both can
benefit from their strengths:
we use abstract interpretation to find loop invariants, an interpolating SMT
solver to analyze single paths, and we combine both in a CEGAR-based abstraction
refinement loop.

\todoin{One paragraph that in the following, we have implemented the approach in
our tool (\automizer) and use that now.}

\subsection{Example}\label{sec:example}
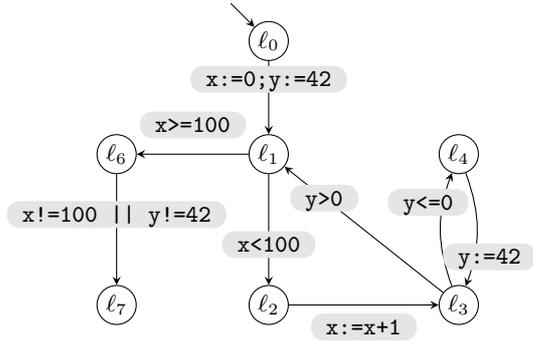
\begin{figure}
\centering
	\begin{subfigure}[b]{0.4\textwidth}
	\centering
	\makebox[0pt]{
\begin{lstlisting}[breaklines=true]
int x:=0, y:=42;
while(x<100){
	x:=x+1;
	while(y<=0){
		y:=42
	}
}
assert x=100 && y=42;

\end{lstlisting}}
	\caption{C code.}
	\label{fig:ex1code}
	\end{subfigure}
	\begin{subfigure}[b]{0.55\textwidth}
	\centering
	\begin{tikzpicture}
\footnotesize

\node[circle,draw,smallnode] (node0) at (0,-1) {$\loc_{0}$};
\node[circle,draw,smallnode] (node1) at (0,-2.5) {$\loc_{1}$};
\node[circle,draw,smallnode] (node2) at (0,-4.5) {$\loc_{2}$};
\node[circle,draw,smallnode] (node3) at (2.5,-4.5) {$\loc_{3}$};
\node[circle,draw,smallnode] (node4) at (2.5,-2.5) {$\loc_{4}$};
\node[circle,draw,smallnode] (node6) at (-2,-2.5) {$\loc_{6}$};
\node[circle,draw,smallnode] (node7) at (-2,-4.5) {$\loc_{7}$};

\draw [trans] ($(-0.5,-0.5)$) to node {} (node0); 

\draw [trans] (node0) to node [near start] 
{\stprog{x:=0;y:=42}} (node1);

\draw [trans] (node1) to node [yshift=-0.2cm]
{\stprog{x<100}} (node2);

\draw [trans] (node2) to node [anchor=north] 
{\stprog{x:=x+1}} (node3);

\draw [trans] (node3) to node [near end]
{\stprog{y>0}} (node1);

\draw [trans, bend left=20] (node3) to node [near end,xshift=-0.2cm]
{\stprog{y<=0}} (node4);

\draw [trans,bend left=20] (node4) to node [near end,xshift=0.2cm] 
{\stprog{y:=42}} (node3);

\draw [trans] (node1) to node [anchor=south]
{\stprog{x>=100}} (node6); 

\draw [trans] (node6) to node [yshift=0.2cm] 
{\stprog{x!=100 || y!=42}} (node7);

\end{tikzpicture}  
	\caption{Control flow graph.}
	\label{fig:ex1cfg}
	\end{subfigure}
\caption{%
Example program \progn{1} with its C code and its corresponding control flow graph (CFG).
The location $\loc_0$ of the CFG is the initial location, $\loc_{7}$ is the
error location.
}
\label{fig:ex1}
\end{figure}

Consider the example program \progn{1} in Figure~\ref{fig:ex1} and its corresponding
control flow graph.
We are interested in proving that the error location of \progn{1}'s control flow
graph (see Figure~\ref{fig:ex1cfg}) $\loc_7$ is unreachable. 
A CEGAR-based approach to generate the proof by iteratively refining an
abstraction of the program
begins with picking a sequence of statements from
the CFG, which starts in the initial location and ends in an error location.
Next, an analysis decides whether the selected sequence of statements is
executable or not, and if not, the abstraction is refined such that this
particular sequence is no longer contained.

Consider the shortest sequence of statements $\tau_1$ from the initial location
$\loc_0$ to the error location $\loc_{7}$.
\[
\tau_1 : \stprog{x:=0;y:=42}\stprog{x>=100}\stprog{x!=100 || y != 42}
\]
This sequence of statements is not executable, because the first two statements
are contradicting each other.
A possible proof for this contradiction consists of the following sequence of
assertions.
\[
\saprog{true}\stprog{x:=0;y:=42}\saprog{x=0}\stprog{x>=100}\saprog{false}
\stprog{x!=100 || y != 42}\saprog{false}
\] 
This sequence of state assertions allows the CEGAR tool to refine its
abstraction such that $\tau_1$ is removed.
In the next iteration, we assume that $\tau_2$ is selected.
\[
\tau_2 : \stprog{x:=0;y:=42}\stprog{x<100}\stprog{x:=x+1}
\stprog{y>0}\stprog{x>=100}\stprog{x!=100 || y != 42}
\]
Again, this sequence of statements is not executable. For example, the
statements \stprog{x:=0;y:=42}, \stprog{x:=x+1} and \stprog{x>=100} contradict
each other, for which we can extract the following proof.
\[
\begin{array}{l}
\saprog{true}\stprog{x:=0;y:=42}
\saprog{x=0}\stprog{x<100}
\saprog{x=0}\stprog{x:=x+1}
\saprog{x=1}\stprog{y>0}\\
\saprog{x=1}\stprog{x>=100}
\saprog{false}\stprog{x!=100 || y != 42}
\saprog{false}
\end{array}
\] 
We can continue in this fashion until we have unrolled the outer while loop of
\progn{1}, but we would rather find other proofs that contain state assertions
that allow us to find a more general refinement of our abstraction, thus
eliminating the need for unrolling.

In our example, the state assertions are not general enough to efficiently
prove the program's correctness, although they were obtained using a
state-of-the-art interpolating SMT solver.
The reason we obtain such assertions is that most solvers prefer to find proofs
for contradictions with as few clauses as possible.\mg{Plz. check this sentence
for sense.}
A more useful but larger reason would involve using the statement \stprog{x!=100
|| y != 42}, where the SMT solver needs to construct a proof of unsatisfiability
that contains both clauses, instead of using \stprog{x >= 100}, where only one
clause has to be contradicted.\mg{Plz. check wording.}
With the help of that larger statement the state assertion \saprog{$x<=100$}
could be obtained.
\saprog{$x<=100$} is very useful because it is a loop invariant at location
$\loc_{1}$, and together with the easily obtained invariant \saprog{$y=42$}, the
two state assertions are sufficient to prove the correctness of our example.

Approaches based on static program analysis, such as abstract
interpretation, can deduce loop invariants by computing a fixpoint for each
program location.
However, such an analysis of the whole program may not be able to find an
invariant strong enough to prove the program to be correct.

\begin{figure}
\centering
\begin{tikzpicture}
\footnotesize

\node[circle,draw,smallnode] 
(node0) at (0,-1) {$\loc_{0}$};

\node[circle,draw,smallnode]
(node1) at (3,-1) {$\loc_{1}$}; 

\node[circle,draw,smallnode]
(node2) at (6,-2) {$\loc_{2}$};

\node[circle,draw,smallnode]
(node3) at (3,-3) {$\loc_{3}$};

\node[circle,draw,smallnode]
(node4) at (6,-1) {$\loc_{6}$};

\node[circle,draw,smallnode] 
(node5) at (10,-1) {$\loc_{7}$};

\draw [trans] ($(-0.75,-0.5)$) to node {} (node0); 

\draw [trans] (node0) to node []
{\stprog{x:=0;y:=42}} (node1);


\draw [trans] (node1) to node []
{\stprog{x<100}} (node2);

%
\draw [trans] (node2) to node []
{\stprog{x:=x+1}} (node3);
%
%
\draw [trans] (node3) to node []
{\stprog{y>0}} (node1);

\draw [trans] (node1) to node []
{\stprog{x>=100}} (node4);

\draw [trans] (node4) to node []
{\stprog{x!=100 || y!=42}} (node5);

\end{tikzpicture}  
\caption{%
The path program computed from the sequence of statements $\tau_2$.}
\label{fig:ex1:pathprogram}
\end{figure}
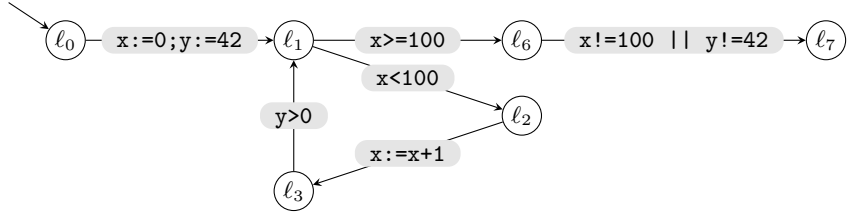

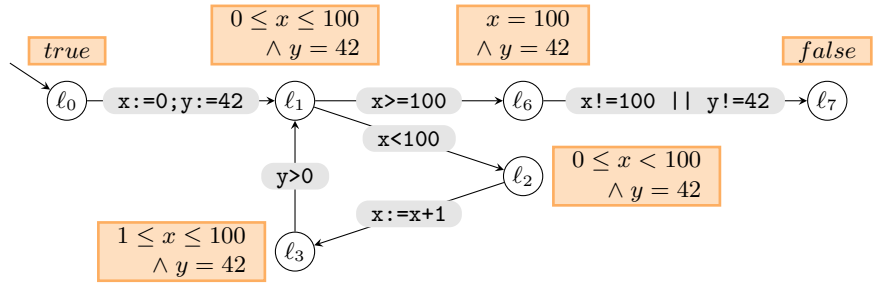
\begin{figure}
\centering


\begin{tikzpicture}
\footnotesize

\node[circle,draw,smallnode,label=above:\saprog{$true$}] 
(node0) at (0,-1) {$\loc_{0}$};

\node[circle,draw,smallnode,label=above:%
\sadyn{$%
\begin{array}{r}%
       0 \leq x \leq 100 \\ 
\wedge \; y=42
\end{array}$}%
]
(node1) at (3,-1) {$\loc_{1}$}; 

\node[circle,draw,smallnode,label=right:%
\sadyn{$%
\begin{array}{r}%
       0 \leq x < 100 \\ 
\wedge \; y=42
\end{array}$}%
]
(node2) at (6,-2) {$\loc_{2}$};

\node[circle,draw,smallnode,label=left:%
\sadyn{$%
\begin{array}{r}%
       1 \leq x \leq 100 \\ 
\wedge \; y=42
\end{array}$}%
]
(node3) at (3,-3) {$\loc_{3}$};

\node[circle,draw,smallnode,label=above:%
\sadyn{$%
\begin{array}{r}%
x=100 \\ 
\wedge \; y=42
\end{array}$}%
]
(node4) at (6,-1) {$\loc_{6}$};

\node[circle,draw,smallnode,label=above:\saprog{$false$}] 
(node5) at (10,-1) {$\loc_{7}$};

\draw [trans] ($(-0.75,-0.5)$) to node {} (node0); 

\draw [trans] (node0) to node []
{\stprog{x:=0;y:=42}} (node1);


\draw [trans] (node1) to node []
{\stprog{x<100}} (node2);

%
\draw [trans] (node2) to node []
{\stprog{x:=x+1}} (node3);
%
%
\draw [trans] (node3) to node []
{\stprog{y>0}} (node1);

\draw [trans] (node1) to node []
{\stprog{x>=100}} (node4);

\draw [trans] (node4) to node []
{\stprog{x!=100 || y!=42}} (node5);

\end{tikzpicture}  
\caption{%
State assertions computed by an interval analysis on the path program from
Figure~\ref{fig:ex1:pathprogram}.}
\label{fig:ex1:intaut}
\end{figure}

One way of improving the precision is not analyzing the whole program but just
a fragment of it.
We can compute such a fragment by projecting the CFG of the program to the
statements from the selected sequence of statements.
The resulting CFG is called a \emph{path
program}~\cite{DBLP:conf/pldi/BeyerHMR07}.
Figure~\ref{fig:ex1:pathprogram} shows the path program computed from \progn{1}
and the sequence of statements $\tau_2$.
We can now calculate the fixpoint of, e.g., an interval abstraction for this
path program, which then yields the state assertions shown in
Figure~\ref{fig:ex1:intaut}.
In this case, the state assertion at $\loc_1$\reviewer{Was $q_1$, changed to
$\loc_1$.} already contains the desired loop
invariant in the second CEGAR iteration.
In general, an interval abstraction may not be enough to find a useful
invariant, but refining the abstraction with conventional methods still allows
the overall algorithm to progress.

In the following, we present our approach that combines the analysis of single
sequences of statements and the analysis of path programs in an
automata-theoretic setting. 
We focus on obtaining loop invariants with abstract interpretation if possible,
but can fall back on the analysis of single traces if the computed abstraction
is too weak to prove infeasibility of a trace.

\section{Preliminaries}\label{sec:prelim}
\reviewerins{The theoretical contribution is weak and the main contribution is
in building a tool that can combine trace refinement and abstract
interpretation. In practice, this is a nice idea but the paper fails to give
convincing arguments that it is working properly.}
In this section, we present our understanding of programs and their semantics,
give a brief overview over abstract interpretation, and explain the trace
abstraction algorithm which we use as basis of our approach.

\subsubsection{Programs and Traces.}
We consider a simple programming language whose statements are assignment,
assume, and sequential composition.
We use the syntax that is defined by the following grammar
\[
 \texttt{s} \ := \ \texttt{assume bexpr} \ \mid\  \texttt{x:=expr}  \ \mid\  \texttt{s;s}
\]
where \var is a finite set of program variables, $\texttt{x} \in \var$,
$\texttt{expr}$ is an expression over $\var$ and $\texttt{bexpr}$ is a Boolean
expression over \var.
For brevity we use \texttt{bexpr} to denote the assume statement \texttt{assume bexpr}.

We represent a \emph{program} over a given set of statements $\Stmt$ as a
labeled graph $\progdef$ with a finite set of nodes $\locs$ called locations, a
set of edges labeled with statements, i.e., $\edgedef$, and a distinguished node
$\loc_{0}$ which we call the initial location.

We call a sequence of statements $\tau=s_0s_1s_2\ldots \in \Stmt^{*}$ a
\emph{trace of the program} $\prog$ if $\tau$ is the edge labeling of a path
that starts at the initial location $\loc_{0}$.
We define the set of all program traces formally as follows:
$$T(\prog) = \{s_0s_1\ldots\in \Stmt^{*} \mid \exists \loc_1,\loc_2,\ldots \bullet
(\loc_i,s_{i},\loc_{i+1}) \in \trans \text{,  for } i \geq 0 \}$$
Note that in each program trace, the source location of the edge labeled with
$s_0$ is the initial location, and therefore, $\loc_{0}$ is not existentially
quantified in the formula for $T(\prog)$.\mg{Added note to take reviewer's
remark into account. Please check.}

Let $\dom$ be the set of values of the program's variables. 
We denote a program state $\sigma$ as a function $\sigma: \var \rightarrow \dom$
that maps program variables to values.
We use $\States$ to denote the set of all program states.
Each statement \texttt{s} $\in\Stmt$ defines a binary relation
$\rho_{\texttt{s}}$ over program states which we call the \emph{successor
relation}.
Let $Expr$ be the set of all expressions over the program variables \var.
We assume a given interpretation function $\ipfname : \mathit{Expr} \times (\var
\rightarrow \dom) \rightarrow \dom$ and define the relation $\rho_{\texttt{s}}
\subseteq \States \times \States$ inductively as follows:
\[
\rho_s =
\begin{cases}
\{ (\sigma,\sigma') \mid \ipf{bexpr}{\sigma} = true \text{ and }\sigma=\sigma' \} 
& \text{ if $s \equiv$ \texttt{assume bexpr}}\\
\{ (\sigma,\sigma') \mid \sigma' = \sigma[\texttt{x} \mapsto \ipf{expr}{\sigma}] \}
& \text{ if $s \equiv$ \texttt{x:=expr}}\\
\{(\sigma,\sigma') \mid \exists \sigma'' \bullet
  (\sigma,\sigma'') \in \rho_{\texttt{s}_1}$ and $(\sigma'',\sigma') \in
  \rho_{\texttt{s}_2} \}
& \text{ if $s \equiv$ \texttt{s$_1$;s$_2$}}\\
\end{cases}
\]

Given a trace $\tau = s_0s_1s_2\ldots$, a sequence of program states $\pi =
\sigma_0\sigma_1\sigma_2\ldots$ is called a \emph{program execution of trace}
$\tau$ if each successive pair of program states is contained in the successor
relation of the corresponding statement of the trace, i.e.,
$(\sigma_{i},\sigma_{i+1}) \in \rho_{s_i}$ for $i\in\{0,1,\ldots\}$.
We call a trace $\tau$ \emph{infeasible} if it does not have any program
execution, otherwise we call $\tau$ \emph{feasible}.
We use $\Pi(\tau)$ to denote the set of all program executions of $\tau$.
The set of all feasible traces of program $\prog$ is denoted by
$\tffeas(\prog)$, and the set of all program executions of $\prog$,
$\Pi(\prog)$, is defined as follows.
$$\Pi(\prog) = \bigcup_{\tau \in \tffeas(\prog)} \Pi(\tau)$$

\subsubsection{Abstract Interpretation}\label{sec:absInt}

Abstract interpretation~\cite{DBLP:conf/popl/CousotC77} is a well-known static
analysis technique that computes a fixpoint of abstract values of an input
program's variables for each program location.
This fixpoint is an over-approximated abstraction of the program's concrete
behavior.
To this end, abstract interpretation uses an \emph{abstract domain} defining
allowed abstract values of the program's variables in the form of a complete
lattice.
The fixpoint computation algorithm analyzes an input program and
annotates each location with an abstract state by iteratively applying
an abstract transformer for each edge, starting at the initial location.
This abstract transformer computes an abstract post state for a given abstract
state and a statement, i.e., it computes the effect a statement has on a given
abstract state.
In case of branching in the program, the fixpoint computation algorithm may
choose to either merge the states at the join point of the branches with a join
operator defined by the abstract domain, or to keep an arbitrary number of
disjunctive states.
In the latter case, precision is increased at the cost of additional
computations due to more abstract states in the abstraction.

The fixpoint computation algorithm is guaranteed to achieve progress and to
eventually terminate, making abstract interpretation one of the most scalable
approaches for program analysis.
Upon termination, an over-approximated abstraction of the program is guaranteed
to have been computed.
Progress is achieved by the application of a widening operator, defined by the
used abstract domain.
When the fixpoint computation algorithm traverses the statements of a loop, an
infinite repetition of the application of the abstract transformer to the
loop's statement is avoided by widening the approximation of the loop's body.
This way, the approximation of the loop is made increasingly wider until a
fixpoint for the effect of the whole loop is found.
\ddins{Too informal. Defining an abstract domain as tuple with lattice,
post, join, meet would help. Use $\rho_s$. Use $\ipfname$. Use program states.}

\subsubsection{Trace Abstraction.}\label{sec:traceabstraction}
The trace abstraction algorithm~\cite{sas/HeizmannHP09, conf/cav/HeizmannHP13}
is a CEGAR-based software model checking approach that proves the correctness of a
program $\prog$ by partitioning the set of possible error traces in feasible and
infeasible traces.
In the following, we briefly explain this approach.   
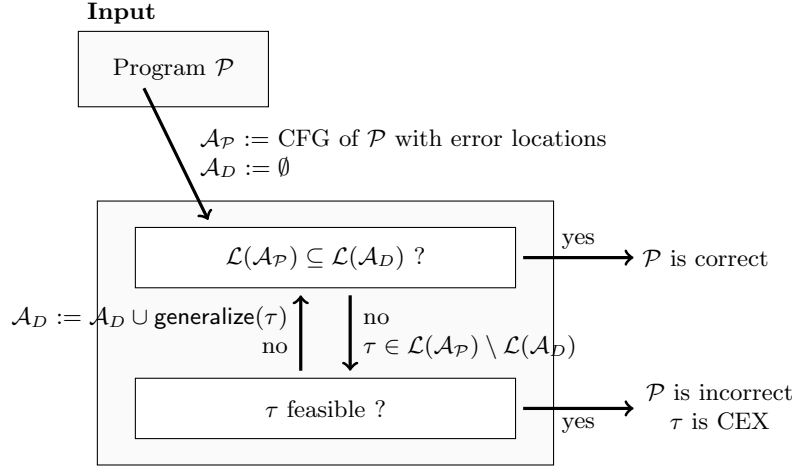
\begin{figure}[t]
\centering
    \settowidth{\imgwidth}{\begin{tikzpicture}[auto]
 
 \node (initP) at (0,6.5) [] {Program \prog};
 
 \node (inclcheck) at (2,4) [check] {$\langatm{\prog} \subseteq \langatm{D}$ ?}; 
 \node (checkfeas) at (2,2) [check] {$\tau$ feasible ?};
 
 \node (safe) at (7,4) [] {\prog is correct}; 
 \node (unsafe) at (7.185,2)[]{%
 \begin{tabular}{c}%
	\prog is incorrect \\
 	$\tau$ is CEX
 \end{tabular}};
 
 \draw [flow,pos=0.6,transform canvas={xshift=-4em}] (initP.south east) to
 node[yshift=-1.75ex,xshift=0.5em] {%
 \begin{tabular}{l}%
	 $\atm{\prog} :=$ CFG of \prog with error locations\\
	 $\atm{D} := \emptyset$
 \end{tabular}} (inclcheck);
 \draw [flow,inner sep=1mm] (inclcheck) to node {yes} (safe);

 \draw [flow,inner sep=1mm,transform canvas={xshift=1em}] (inclcheck) to node {%
 \begin{tabular}{l}%
 	no\\
	$\tau \in \langatm{\prog} \setminus \langatm{D}$
 \end{tabular}} (checkfeas);

 \draw [flow,inner sep=1mm,transform canvas={xshift=-1em}] (checkfeas) to node {%
 \begin{tabular}{r}%
 	$\atm{D} := \atm{D} \cup \refF(\tau)$\\
 	no
 \end{tabular}} (inclcheck);

\draw[flow,inner sep=1mm,swap] (checkfeas.east) -- node{yes} (unsafe);

\begin{pgfonlayer}{background}
 \draw [fill=lightback] (-1.25,6.0) rectangle (1.25,7.0);
 \node at (-1.25,7.5) [below right] {\textbf{Input}};
 \draw [fill=lightback] (-1,1.25) rectangle (5,4.75);
\end{pgfonlayer}

\end{tikzpicture}}%
    \setlength{\imgwidth}{\minof{\imgwidth}{\textwidth}}%
    \resizebox{\imgwidth}{!}{\begin{tikzpicture}[auto]
 
 \node (initP) at (0,6.5) [] {Program \prog};
 
 \node (inclcheck) at (2,4) [check] {$\langatm{\prog} \subseteq \langatm{D}$ ?}; 
 \node (checkfeas) at (2,2) [check] {$\tau$ feasible ?};
 
 \node (safe) at (7,4) [] {\prog is correct}; 
 \node (unsafe) at (7.185,2)[]{%
 \begin{tabular}{c}%
	\prog is incorrect \\
 	$\tau$ is CEX
 \end{tabular}};
 
 \draw [flow,pos=0.6,transform canvas={xshift=-4em}] (initP.south east) to
 node[yshift=-1.75ex,xshift=0.5em] {%
 \begin{tabular}{l}%
	 $\atm{\prog} :=$ CFG of \prog with error locations\\
	 $\atm{D} := \emptyset$
 \end{tabular}} (inclcheck);
 \draw [flow,inner sep=1mm] (inclcheck) to node {yes} (safe);

 \draw [flow,inner sep=1mm,transform canvas={xshift=1em}] (inclcheck) to node {%
 \begin{tabular}{l}%
 	no\\
	$\tau \in \langatm{\prog} \setminus \langatm{D}$
 \end{tabular}} (checkfeas);

 \draw [flow,inner sep=1mm,transform canvas={xshift=-1em}] (checkfeas) to node {%
 \begin{tabular}{r}%
 	$\atm{D} := \atm{D} \cup \refF(\tau)$\\
 	no
 \end{tabular}} (inclcheck);

\draw[flow,inner sep=1mm,swap] (checkfeas.east) -- node{yes} (unsafe);

\begin{pgfonlayer}{background}
 \draw [fill=lightback] (-1.25,6.0) rectangle (1.25,7.0);
 \node at (-1.25,7.5) [below right] {\textbf{Input}};
 \draw [fill=lightback] (-1,1.25) rectangle (5,4.75);
\end{pgfonlayer}

\end{tikzpicture}}%

\caption{The trace abstraction algorithm.}
\label{fig:ta:algo}
\end{figure}
Consider the trace abstraction algorithm shown in Figure~\ref{fig:ta:algo}.
The input program $\prog$ over the set of statements \Stmt is first translated
into a \emph{program automaton} $\atm{\prog}$, which encodes the correctness
property of $\prog$ by marking some of its locations as error locations.
Those error locations serve as the accepting states of the program automaton
$\atm{\prog}$, and the set of statements \Stmt as its alphabet $\Sigma$.
By construction, every word accepted by this automaton represents a trace of
$\prog$ that can reach the error location.
Next, the algorithm determines whether the language of \atm{\prog} contains a
feasible trace, which would then be a valid counterexample.
To this end, a \emph{data automaton} $\atm{D}$ over the same alphabet as
\atm{\prog} is constructed such that its language consists only of infeasible
traces.
More formally, a data automaton is a Floyd-Hoare
automaton~\cite{conf/cav/HeizmannHP13, phddietsch}.
A Floyd-Hoare automaton $\atm{}=(Q,\delta,q_0,F)$ is an automaton over the
alphabet of the program's statements $\Stmt$ together with a mapping that
assigns to each state $q\in Q$ a formula $\varphi_q$ that denotes a predicate
over the program variables such that the following holds:
\begin{itemize}[itemsep=0pt,topsep=0pt]
 \item The initial state is annotated by the formula $\mathit{true}$.
 \item For each transition $(q, \mathit{st}, q') \in \delta$ the
 triple $\hoare{\varphi_q}{\mathit{st}}{\varphi_{q'}}$ is a valid Hoare
 triple.
 \item Each accepting state $q \in F$ is annotated by the formula
 $\mathit{false}$.
\end{itemize}
Initially, the data automaton $\atm{D}$ is empty.   
In each iteration, the algorithm checks whether the language of the current data
automaton \atm{D} is a superset of the language of the program automaton
\atm{\prog}.
If this is the case, all traces in \atm{\prog} are infeasible, i.e., the error
locations of program \prog cannot be reached.
If this is not the case, there exists a trace $\tau$ of
\atm{\prog} which is not in \atm{D}, and thus not known to be infeasible. 

Therefore, if the trace $\tau$ is feasible, it represents at least one valid
program execution that can reach an error location. 
If the trace $\tau$ is infeasible, the algorithm constructs a new data automaton
\atm{D} whose language contains more infeasible traces than the old by computing
a union of the old automaton \atm{D} and a new automaton obtained by
generalizing the proof of infeasibility of $\tau$ (\refF).\reviewer{Where is
\refF defined?}\mg{Define the generalization of the automaton\ldots}

\section{Algorithm}\label{sec:algo}

%
%

\begin{figure}[t]
\centering
    \settowidth{\imgwidth}{\begin{tikzpicture}[auto]
 
 \node (initP) at (0,6.5) [] {Program \prog};
 
 \node (inclcheck) at (2,4) [check] {$\langatm{\prog} \subseteq \langatm{D}$ ?}; 
 \node (checkfeas) at (2,2) [check] {$\tau$ feasible ?};
 \node (checkabs) at (2,0) [check] {AI provides proof ?};
 
 \node (safe) at (6.85,4) [] {%
 \begin{tabular}{c}
 	\prog is correct
 \end{tabular}};
 
 \node (unsafe) at (7,2) [] {%
 \begin{tabular}{c}%
	\prog is incorrect \\
 	$\tau$ is CEX
 \end{tabular}};
 
 \draw [flow,pos=0.6,transform canvas={xshift=-4em}] (initP.south east) to
 node[yshift=-1ex, xshift=0.5em] {%
 \begin{tabular}{l}%
	 $\atm{\prog} :=$ CFG of $\prog$ with error locations\\
	 $\atm{D} := \emptyset$
 \end{tabular}} (inclcheck);
 \draw [flow,inner sep=1mm] (inclcheck) to node {yes} (safe);

 \draw [flow,inner sep=1mm] (inclcheck) to node {%
 \begin{tabular}{l}%
 	no\\
	$\tau \in \langatm{\prog} \setminus \langatm{D}$
 \end{tabular}} (checkfeas);

\draw [flow, inner sep=1mm] (checkfeas) to node {%
\begin{tabular}{l}%
	no\\
	$\textsf{absInt}(\textsf{pathProg}(\tau))\xspace$%
\end{tabular}} (checkabs);

\draw [flow,inner sep=1mm, -, transform canvas={yshift=1ex}] (checkabs) --
(-1.25,0);
\draw [flow,inner sep=1mm, -] ([yshift=1ex]-1.25,0) -- node
{%
  \begin{tabular}{r}%
  	$\atm{D} := \atm{D} \cup \refF(\tau)$\\
  	no
  \end{tabular}} ([yshift=-1ex]-1.25,4);
\draw [flow,inner sep=1mm, transform canvas={yshift=-1ex}] (-1.25,4) --
(inclcheck);

\draw [flow, inner sep=1mm, -, transform canvas={yshift=-1ex}]
(checkabs) -- (-5.75,0);
\draw [flow, inner sep=1mm, -] ([yshift=-1ex]-5.75,0) -- node
[right, yshift=-5em] {%
	\begin{tabular}{l}%
		$\atm{D} := \atm{D} \cup \atm{\mathit{AI}}$\\
		yes
	\end{tabular}} ([yshift=1ex]-5.75,4);
\draw [flow, inner sep=1mm, transform canvas={yshift=1ex}] (-5.75,4) --
(inclcheck);


\draw[flow,inner sep=1mm,swap] (checkfeas.east) -- node{yes} (unsafe);

\begin{pgfonlayer}{background}
 \draw [fill=lightback] (-1.25,6.0) rectangle (1.25,7.0);
 \node at (-1.25,7.5) [below right] {\textbf{Input}};
 \draw [fill=lightback] (-1,-0.75) rectangle (5,4.75);
\end{pgfonlayer}

\end{tikzpicture}}%
    \setlength{\imgwidth}{\minof{\imgwidth}{\textwidth}}%
    \resizebox{\imgwidth}{!}{\begin{tikzpicture}[auto]
 
 \node (initP) at (0,6.5) [] {Program \prog};
 
 \node (inclcheck) at (2,4) [check] {$\langatm{\prog} \subseteq \langatm{D}$ ?}; 
 \node (checkfeas) at (2,2) [check] {$\tau$ feasible ?};
 \node (checkabs) at (2,0) [check] {AI provides proof ?};
 
 \node (safe) at (6.85,4) [] {%
 \begin{tabular}{c}
 	\prog is correct
 \end{tabular}};
 
 \node (unsafe) at (7,2) [] {%
 \begin{tabular}{c}%
	\prog is incorrect \\
 	$\tau$ is CEX
 \end{tabular}};
 
 \draw [flow,pos=0.6,transform canvas={xshift=-4em}] (initP.south east) to
 node[yshift=-1ex, xshift=0.5em] {%
 \begin{tabular}{l}%
	 $\atm{\prog} :=$ CFG of $\prog$ with error locations\\
	 $\atm{D} := \emptyset$
 \end{tabular}} (inclcheck);
 \draw [flow,inner sep=1mm] (inclcheck) to node {yes} (safe);

 \draw [flow,inner sep=1mm] (inclcheck) to node {%
 \begin{tabular}{l}%
 	no\\
	$\tau \in \langatm{\prog} \setminus \langatm{D}$
 \end{tabular}} (checkfeas);

\draw [flow, inner sep=1mm] (checkfeas) to node {%
\begin{tabular}{l}%
	no\\
	$\textsf{absInt}(\textsf{pathProg}(\tau))\xspace$%
\end{tabular}} (checkabs);

\draw [flow,inner sep=1mm, -, transform canvas={yshift=1ex}] (checkabs) --
(-1.25,0);
\draw [flow,inner sep=1mm, -] ([yshift=1ex]-1.25,0) -- node
{%
  \begin{tabular}{r}%
  	$\atm{D} := \atm{D} \cup \refF(\tau)$\\
  	no
  \end{tabular}} ([yshift=-1ex]-1.25,4);
\draw [flow,inner sep=1mm, transform canvas={yshift=-1ex}] (-1.25,4) --
(inclcheck);

\draw [flow, inner sep=1mm, -, transform canvas={yshift=-1ex}]
(checkabs) -- (-5.75,0);
\draw [flow, inner sep=1mm, -] ([yshift=-1ex]-5.75,0) -- node
[right, yshift=-5em] {%
	\begin{tabular}{l}%
		$\atm{D} := \atm{D} \cup \atm{\mathit{AI}}$\\
		yes
	\end{tabular}} ([yshift=1ex]-5.75,4);
\draw [flow, inner sep=1mm, transform canvas={yshift=1ex}] (-5.75,4) --
(inclcheck);


\draw[flow,inner sep=1mm,swap] (checkfeas.east) -- node{yes} (unsafe);

\begin{pgfonlayer}{background}
 \draw [fill=lightback] (-1.25,6.0) rectangle (1.25,7.0);
 \node at (-1.25,7.5) [below right] {\textbf{Input}};
 \draw [fill=lightback] (-1,-0.75) rectangle (5,4.75);
\end{pgfonlayer}

\end{tikzpicture}}%

\caption{The trace abstraction algorithm with abstract interpretation
refinement.}
\label{fig:cegarAi}
\end{figure}
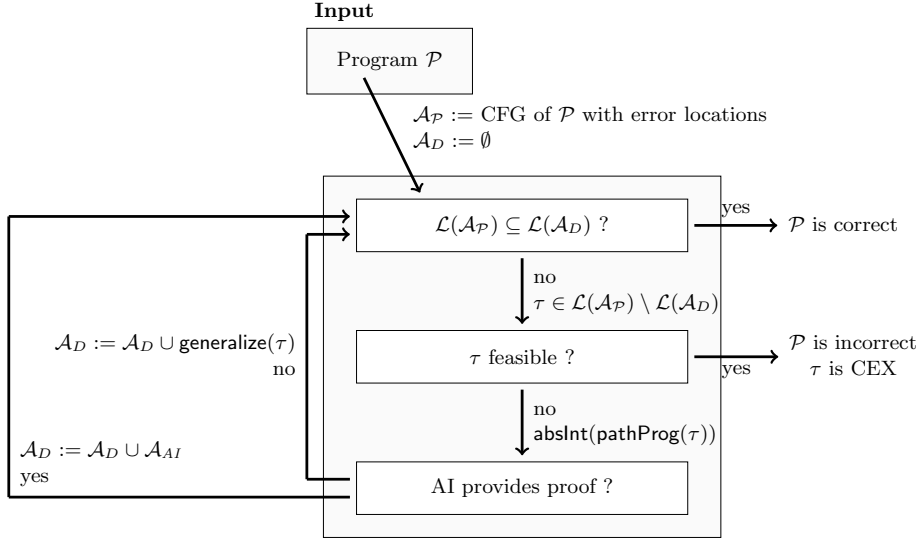

In this section we present a modified version of the trace abstraction CEGAR
loop introduced in Section~\ref{sec:traceabstraction}, which uses a new method
based on abstract interpretation for obtaining the data automaton.
Our algorithm is shown in Figure~\ref{fig:cegarAi}. 
As in the default trace abstraction algorithm, the input program $\prog$ is
translated into a program automaton $\atm{\prog}$.
The initial data automaton $\atm{D}$ is empty.

After analyzing a possible counterexample trace $\tau$ for infeasibility, we
first construct a \emph{path program} $\pprogtau$ from the trace.
We use the theoretical foundations of path programs provided by Beyer et
al.~\cite{DBLP:conf/pldi/BeyerHMR07}.
In our context, a path program is defined as follows.

A path program $\pprog$ of program $\progdef$ with statements $\Stmt$ and
trace $\tau = s_0 s_1 \ldots \in 2^{\Stmt}$ of $\prog$ is a program $\pprog =
(\locs^{\#}, \delta^{\#}, \loc_0^{\#})$ such that
\begin{itemize}[itemsep=0pt,topsep=0pt]
  \item the set of program locations $\locs^{\#}$ contains only locations that
  are visited by trace $\tau$, i.e.,
  $\locs^{\#} = \{ \loc \mid \loc \in \locs \wedge \exists s_i \in
  \tau \; s.t.\; (\loc,s_i,\loc') \in \delta \vee (\loc',s_i,\loc) \in
  \delta)\}$,
  \item the transition relation $\delta^{\#}$ contains only transitions labeled
  with symbols from trace $\tau$, i.e., $\delta^{\#} = \{ (\loc,s_i,\loc') \mid
  s_i \in \tau \wedge (\loc,s_i,\loc') \in \delta \}$, and 
  \item the initial location $\loc_0^{\#}$ stays the same, i.e., $\loc_0^{\#} =
  \loc_0$.
\end{itemize}

After constructing a path program of $\tau$, \pprogtau, we compute an
abstraction of \pprogtau using abstract interpretation.
%
%
If the abstraction provides a proof for the infeasibility of the path program,
we have obtained suitable loop invariants for all loops that occur in the path
program.
In this case we construct a data automaton $\atm{\mathit{AI}}$ from $\pprogtau$
which is then added to the existing data automaton $\atm{D}$.
In the case where abstract interpretation fails to prove infeasibility of the
path program, we generalize the trace $\tau$ with the generalization method
from the trace abstraction algorithm.
Therefore, our approach is able to retain the precision of trace abstraction,
but has a useful mechanism to prevent divergence due to loop unrolling.
In the best case, we converge faster than the trace abstraction algorithm
because we are able to find suitable loop invariants for the investigated
path programs.

\medskip

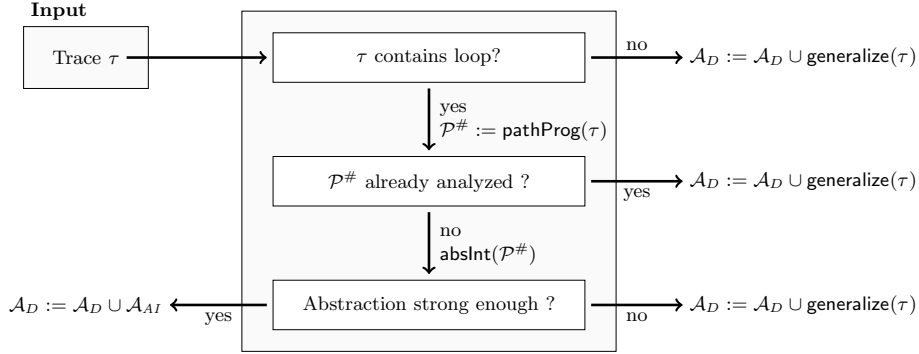
\begin{figure}[t]
\centering
    \settowidth{\imgwidth}{\begin{tikzpicture}[auto]
 
 \node (initP) at (-3.5,4.0) [] {Trace $\tau$};
 
 \node (loopcheck) at (2,4) [check] {$\tau$ contains loop?}; 
 \node (ppana) at (2,2) [check] {$\pprog$ already analyzed ?};
 \node (checkabs) at (2,0) [check] {Abstraction strong enough ?};
 
 \node (defta1) at (8,4) [] {$\atm{D} := \atm{D} \cup \refF(\tau)$};
 \node (defta2) at (8,2) [] {$\atm{D} := \atm{D} \cup \refF(\tau)$};
 \node (defta3) at (8,0) [] {$\atm{D} := \atm{D} \cup \refF(\tau)$};

 \node (absref) at (-3.5,0) [] {$\atm{D} := \atm{D} \cup \atm{\mathit{AI}}$};
 \draw [flow,pos=0.6] (initP) to (inclcheck);
\draw [flow,inner sep=1mm] (inclcheck) to node {no} (defta1);

 \draw [flow,inner sep=1mm] (inclcheck) to node {%
 \begin{tabular}{l}%
 	yes\\
 	$\pprog := \textsf{pathProg}(\tau)$
 \end{tabular}} (ppana);

\draw[flow,inner sep=1mm,swap] (ppana.east) -- node{yes} (defta2);

\draw[flow,inner sep=1mm] (ppana) -- node {%
\begin{tabular}{l}%
no\\
$\textsf{absInt}(\pprog)$
\end{tabular}} (checkabs);

\draw[flow,inner sep=1mm,swap] (checkabs.east) -- node{no} (defta3);

\draw[flow,inner sep=1mm] (checkabs.west) -- node{yes} (absref);

\begin{pgfonlayer}{background}
 \draw [fill=lightback] (-4.5,3.5) rectangle (-2.5,4.5);
 \node at (-4.5,5.0) [below right] {\textbf{Input}};
 \draw [fill=lightback] (-1,-0.75) rectangle (5,4.75);
\end{pgfonlayer}

\end{tikzpicture}}%
    \setlength{\imgwidth}{\minof{\imgwidth}{\textwidth}}%
    \resizebox{\imgwidth}{!}{\begin{tikzpicture}[auto]
 
 \node (initP) at (-3.5,4.0) [] {Trace $\tau$};
 
 \node (loopcheck) at (2,4) [check] {$\tau$ contains loop?}; 
 \node (ppana) at (2,2) [check] {$\pprog$ already analyzed ?};
 \node (checkabs) at (2,0) [check] {Abstraction strong enough ?};
 
 \node (defta1) at (8,4) [] {$\atm{D} := \atm{D} \cup \refF(\tau)$};
 \node (defta2) at (8,2) [] {$\atm{D} := \atm{D} \cup \refF(\tau)$};
 \node (defta3) at (8,0) [] {$\atm{D} := \atm{D} \cup \refF(\tau)$};

 \node (absref) at (-3.5,0) [] {$\atm{D} := \atm{D} \cup \atm{\mathit{AI}}$};
 \draw [flow,pos=0.6] (initP) to (inclcheck);
\draw [flow,inner sep=1mm] (inclcheck) to node {no} (defta1);

 \draw [flow,inner sep=1mm] (inclcheck) to node {%
 \begin{tabular}{l}%
 	yes\\
 	$\pprog := \textsf{pathProg}(\tau)$
 \end{tabular}} (ppana);

\draw[flow,inner sep=1mm,swap] (ppana.east) -- node{yes} (defta2);

\draw[flow,inner sep=1mm] (ppana) -- node {%
\begin{tabular}{l}%
no\\
$\textsf{absInt}(\pprog)$
\end{tabular}} (checkabs);

\draw[flow,inner sep=1mm,swap] (checkabs.east) -- node{no} (defta3);

\draw[flow,inner sep=1mm] (checkabs.west) -- node{yes} (absref);

\begin{pgfonlayer}{background}
 \draw [fill=lightback] (-4.5,3.5) rectangle (-2.5,4.5);
 \node at (-4.5,5.0) [below right] {\textbf{Input}};
 \draw [fill=lightback] (-1,-0.75) rectangle (5,4.75);
\end{pgfonlayer}

\end{tikzpicture}}%

\caption{Abstract interpretation module.}
\label{fig:aimodule}
\end{figure}

In the following, we describe the abstract interpretation module of the
algorithm in Figure~\ref{fig:cegarAi} in more detail.
The basic functionality of our abstract interpretation module is depicted in
Figure~\ref{fig:aimodule}.
The abstract interpretation module is used when a trace $\tau$ has been
identified as being infeasible.
First, it is determined whether $\tau$ contains a loop.
If $\tau$ does not contain a loop, trace abstraction's generalization can
easily compute state assertions with the help of an SMT solver as no loop
invariants are needed.
If $\tau$ contains at least one loop, i.e., if there exists a statement in
$\tau$ which is part of a loop in $\prog$, we use $\tau$ to construct a path
program, $\pprogtau$.

Next, we check whether $\pprogtau$ corresponds to a path program which has
already been analyzed in a previous CEGAR iteration to avoid analyzing the same
path program twice.
This may happen if abstract interpretation was unable to find a proof for a
path program in an earlier iteration and we are in the process of unrolling the
loop.
In this case, we have to continue with the standard trace abstraction refinement
step to ensure progress.
Otherwise, we use abstract interpretation to compute a fixpoint abstraction of
$\pprogtau$.

Abstract interpretation can yield two possible results: $\pprogtau$ is proven to be
safe, or the computed abstraction is too weak to prove safety.
If $\pprogtau$ is proven to be safe, i.e., the error location of $\pprogtau$ is
unreachable, the state assertions obtained through the computed abstraction are
a proof for the trace's infeasibility.
In this case, we construct a data automaton $\atm{\mathit{AI}}$ from the
generated state assertions which is then added to the existing data automaton
$\atm{D}$.
If the error location of $\pprogtau$ is reachable because of a too coarse
abstraction, we use trace abstraction's generalization method to generalize
$\tau$ and continue with the next CEGAR iteration.

\subsection{Data Automaton Construction from Path Programs}
We construct a data automaton $\atm{\mathit{AI}} = (Q,\delta_{\mathit{AI}}, q_0,
F)$ from a path program $\pprogtau = (\locs^{\#}, \delta^{\#}, \loc_0^{\#})$,
annotated with state assertions obtained through the computed abstraction as
follows.
\begin{itemize}[itemsep=0pt,topsep=0pt]
  \item The set of locations $Q$ of $\atm{\mathit{AI}}$ contains a location
  for each unique fixpoint computed for $\pprog$, i.e., \\
  $\forall q_1, q_2 \in Q \; \exists \loc^{\#}_1, \loc^{\#}_2 \in \locs^{\#}
  \; s.t. \; \varphi_{\loc^{\#}_1} = \varphi_{q_1} \wedge \varphi_{\loc^{\#}_2}
  = \varphi_{q_2} \wedge \varphi_{q_1} = \varphi_{q_2} \iff q_1 = q_2$,
  \item the set of transitions of $\atm{\mathit{AI}}$ corresponds to the set of
  transitions of $\pprog$ with respect to the locations in $Q$, i.e.,
  $\forall (q, \mathit{st}, q') \in \delta_{\mathit{AI}} \; \exists
  \loc^{\#}_1, \loc^{\#}_2 \in \locs^{\#} \; s.t. \; \varphi_{q} =
  \varphi_{\loc^{\#}_1} \wedge \varphi_{q'} = \varphi_{\loc^{\#}_2}
  \wedge (\loc^{\#}_1 , \mathit{st} , \loc^{\#}_2 ) \in \delta^{\#}$,
  \item the initial location of $\atm{\mathit{AI}}$ and $\pprog$ are the same,
  i.e., $q_0 = \loc_0^{\#}$,
  \item the set of accepting states $F$ contains the error location of
  $\pprogtau$, and
  \item every state $q \in Q$ is annotated with a formula $\varphi_q$ which is
  the state assertion computed by the fixpoint engine of the path
  program location corresponding to $q$.
\end{itemize}
By construction, we retain the property of Floyd-Hoare automata, that for each
transition $(q, \mathit{st}, q') \in \delta_\mathit{AI}$ the triple
$\hoare{\varphi_q}{\mathit{st}}{\varphi_{q'}}$ is a valid Hoare triple.
Therefore, the automaton accepts at least all the traces represented by the path
program.

\begin{figure}[t]
\centering
\resizebox{0.95\textwidth}{!}{


\begin{tikzpicture}
\footnotesize

\node[circle,draw,smallnode,label=above:\saprog{$true$}] 
(node0) at (0,-1) {$\loc_{0}$};

\node[circle,draw,smallnode,label=above:%
\sadyn{$%
\begin{array}{r}%
       0 \leq x \leq 100 \\ 
\wedge \; y=42
\end{array}$}%
]
(node1) at (3,-1) {$\loc_{1}$}; 

\node[circle,draw,smallnode,label=below:%
\sadyn{$%
\begin{array}{r}%
       0 \leq x \leq 99 \\ 
\wedge \; y=42
\end{array}$}%
] (node2) at (7,-4) {$\loc_{2}$};

\node[circle,draw,smallnode,label=below:%
\sadyn{$%
\begin{array}{r}%
       1 \leq x \leq 100 \\ 
\wedge \; y=42
\end{array}$}%
]
(node3) at (3,-4) {$\loc_{3}$};

\node[circle,draw,smallnode,label=above:%
\sadyn{$%
\begin{array}{r}%
x=100 \\ 
\wedge \; y=42
\end{array}$}%
] (node6) at (7,-1) {$\loc_{6}$};

\node[circle,draw,smallnode,accepting, label=above:\saprog{$false$}] (node7) 
at (11.5,-1) {$\loc_{7}$};

\draw [trans] ($(-0.75,-0.5)$) to node {} (node0); 

\draw [trans] (node0) to node []
{\stprog{x:=0;y:=42}} (node1);

\draw [trans,bend left=20] (node1) to node [near end]
{\stprog{x<100}} (node2);

\draw [trans] (node2) to node [near end,below]
{$\Sigma$} (node1);

\draw [trans] (node2) to node [yshift=0.2cm]
{$\Sigma \setminus \{ \stprog{x:=0;y:=42} \}$} (node3);

\draw[trans,bend right=20] (node3) to node[] 
{\stprog{x<100}}
(node2);

\draw [trans,left,bend left=20] (node3) to node []
{$\Sigma \setminus \{ \stprog{x:=x+1} \} $ } 
(node1);

\draw [trans,right] (node1) to node []
{\rotatebox{270}{$\Sigma \setminus \{ \stprog{x:=x+1} \} $} } 
(node3);

\draw [trans,bend left=20] (node1) to node []
{\stprog{x>=100}} (node6);

\draw [trans] (node6) to node [yshift=-0.2cm,xshift=0.2cm]
{$\Sigma \setminus \{ \stprog{x:=x+1} \}$} (node1);

\draw [trans] (node6) to node []
{\stprog{x!=100 || y!=42}} (node7);

\draw [trans] (node3) to node [near start,xshift=0.2cm]
{\stprog{x>=100}} (node6);

\draw [trans,out=235,in=205,looseness=12] (node0) to 
node [yshift=-0.2cm,xshift=-0.2cm] 
{$\Sigma$} (node0);
 
\draw [trans,out=235,in=205,looseness=12] (node1) to 
node[yshift=-0.25cm,xshift=-1.1cm] 
{$\Sigma \setminus \{ \stprog{x:=x+1} \} $ }
(node1);

\draw [trans,loop right] (node2) to 
node []
{$\Sigma \setminus \{ \stprog{x:=x+1} \} $} (node2);

\draw [trans,loop left] (node3) to node []
{$\Sigma \setminus \{
\begin{array}{c}
\stprog{x:=x+1}\\
\stprog{x:=0;y:=42}
\end{array}\}
$} (node3);

\draw [trans,out=335,in=305,looseness=12] (node6) to node 
[yshift=-0.25cm,xshift=1.5cm]
{$\Sigma \setminus \{
\begin{array}{c}
\stprog{x:=x+1}\\
\stprog{x:=0;y:=42}
\end{array}\}
$} (node6);

\draw [trans,loop below] (node7) to node [] 
{$\Sigma$} (node7);

\end{tikzpicture}  }
\caption{%
Enhanced data automaton computed by an interval analysis on the path
program from Figure~\ref{fig:ex1:pathprogram} with $%
\Sigma = \{ \stprog{x:=0;y:=42}, \stprog{x<100}, \stprog{x:=x+1}, \stprog{y>0},$
$\stprog{x>=100}, \stprog{x!=100 || y!=42}, \stprog{y<=0}, \stprog{y:=42} \}$.}
\label{fig:ex1:intaut:pp:total}
\end{figure}
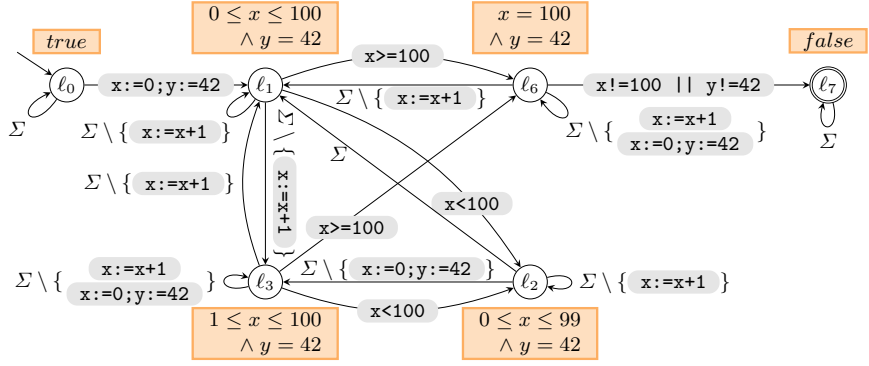

The constructed data automaton can be further enhanced to exclude more traces
by generalizing it as follows.
For each triple $\hoare{\varphi_{q}}{\mathit{st}}{\varphi_{q'}}$, where $q,
q',\mathit{st} \in \atm{\mathit{AI}}$, not already represented in
$\atm{\mathit{AI}}$ it is checked whether the triple is a valid Hoare
triple.
We do that by computing the abstract post state of the abstract state of $q$ and
the statement $\mathit{st}$.
If the resulting post state is a subset of $q'$, we have found a valid Hoare
triple.
Otherwise, the triple is not a valid Hoare triple with respect to the current
abstraction.
For all of those valid Hoare triples, a new transition between the corresponding
states, labeled with the corresponding statements is added to
$\atm{\mathit{AI}}$.
Figure~\ref{fig:ex1:intaut:pp:total} shows the enhanced data automaton for the
example path program from Section~\ref{sec:example}.

\section{Implementation and Evaluation}\label{sec:impl}
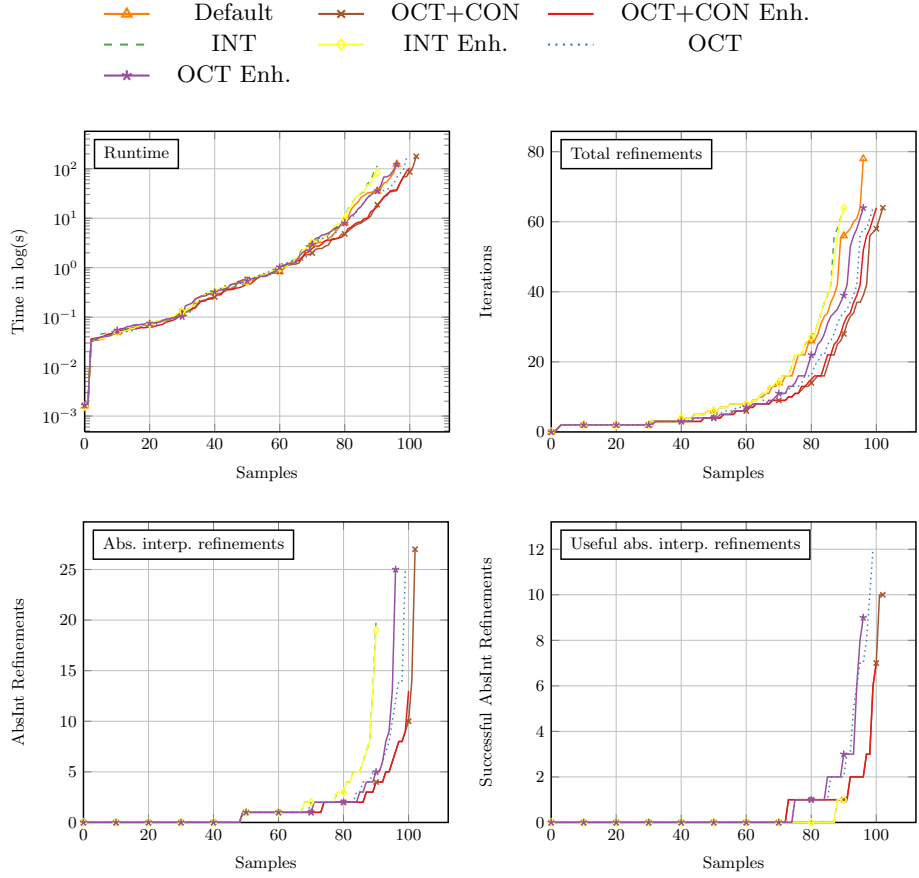
\begin{figure}[!t]
\centering
\def\plotscale{0.7}
	\begin{subfigure}{\textwidth}
	\centering
	\begin{tikzpicture}
\begin{customlegend}[legend columns=3,legend style={align=left,draw=none,column sep=2ex,thick},
                      legend entries={\setAuto,\setComp,\setCompT,\setInt,\setIntT,\setOct,\setOctT,}]
    \addlegendimage{mark repeat={10},draw=s5,solid,mark=triangle}
    \addlegendimage{mark repeat={10},draw=s7,solid,mark=x}
    \addlegendimage{draw=s1,solid}
    \addlegendimage{draw=s3,dashed}
    \addlegendimage{mark repeat={10},draw=s6,solid,mark=diamond}
    \addlegendimage{draw=s2,dotted}
    \addlegendimage{mark repeat={10},draw=s4,solid,mark=star}
\end{customlegend}
\end{tikzpicture}
	\phantomcaption
	\end{subfigure}
	\par\bigskip
	\begin{subfigure}[t]{0.5\textwidth}
	\centering
	\begin{tikzpicture}[scale=\plotscale]
\begin{semilogyaxis}[%
xmin=0, ymin=0,%
xlabel={Samples},%
ylabel={Time in log(s)},grid=major,%
legend style={at={(0.025,0.975)},anchor=north west,legend cell align=left}%
]%
\addlegendimage{empty legend}\addlegendentry{Runtime}
\addplot[mark repeat={10},draw=s5,solid,mark=triangle] table {plots/Runtime-svcomp-Reach-32bit-Automizer_Default.epf.plot};
\addplot[mark repeat={10},draw=s7,solid,mark=x] table {plots/Runtime-svcomp-Reach-32bit-Automizer_Default+AIv2_COMP_Simple.epf.plot};
\addplot[draw=s1,solid] table {plots/Runtime-svcomp-Reach-32bit-Automizer_Default+AIv2_COMP_Simple_total.epf.plot};
\addplot[draw=s3,dashed] table {plots/Runtime-svcomp-Reach-32bit-Automizer_Default+AIv2_INT.epf.plot};
\addplot[mark repeat={10},draw=s6,solid,mark=diamond] table {plots/Runtime-svcomp-Reach-32bit-Automizer_Default+AIv2_INT_total.epf.plot};
\addplot[draw=s2,dotted] table {plots/Runtime-svcomp-Reach-32bit-Automizer_Default+AIv2_OCT.epf.plot};
\addplot[mark repeat={10},draw=s4,solid,mark=star] table {plots/Runtime-svcomp-Reach-32bit-Automizer_Default+AIv2_OCT_total.epf.plot};
\end{semilogyaxis}
\end{tikzpicture}
	\phantomcaption
	\end{subfigure}
	\hfil%
	\begin{subfigure}[t]{0.49\textwidth}
	\centering
	\begin{tikzpicture}[scale=\plotscale]
\begin{axis}[%
xmin=0, ymin=0,%
xlabel={Samples},%
ylabel={Iterations},grid=major,%
legend style={at={(0.025,0.975)},anchor=north west,legend cell align=left}%
]%
\addlegendimage{empty legend}\addlegendentry{Total refinements}
\addplot[mark repeat={10},draw=s5,solid,mark=triangle] table {plots/TotalIterations-svcomp-Reach-32bit-Automizer_Default.epf.plot};
\addplot[mark repeat={10},draw=s7,solid,mark=x] table {plots/TotalIterations-svcomp-Reach-32bit-Automizer_Default+AIv2_COMP_Simple.epf.plot};
\addplot[draw=s1,solid] table {plots/TotalIterations-svcomp-Reach-32bit-Automizer_Default+AIv2_COMP_Simple_total.epf.plot};
\addplot[draw=s3,dashed] table {plots/TotalIterations-svcomp-Reach-32bit-Automizer_Default+AIv2_INT.epf.plot};
\addplot[mark repeat={10},draw=s6,solid,mark=diamond] table {plots/TotalIterations-svcomp-Reach-32bit-Automizer_Default+AIv2_INT_total.epf.plot};
\addplot[draw=s2,dotted] table {plots/TotalIterations-svcomp-Reach-32bit-Automizer_Default+AIv2_OCT.epf.plot};
\addplot[mark repeat={10},draw=s4,solid,mark=star] table {plots/TotalIterations-svcomp-Reach-32bit-Automizer_Default+AIv2_OCT_total.epf.plot};
\end{axis}
\end{tikzpicture}
	\phantomcaption
	\end{subfigure}	
	\par\bigskip
	\begin{subfigure}[t]{0.5\textwidth}
	\centering
	\begin{tikzpicture}[scale=\plotscale]
\begin{axis}[%
xmin=0, ymin=0,%
xlabel={Samples},%
ylabel={AbsInt Refinements},grid=major,%
legend style={at={(0.025,0.975)},anchor=north west,legend cell align=left}%
]%
\addlegendimage{empty legend}
\addlegendentry{Abs. interp. refinements} 
\addplot[mark repeat={10},draw=s7,solid,mark=x] table
{plots/AIIterations-svcomp-Reach-32bit-Automizer_Default+AIv2_COMP_Simple.epf.plot}; \addplot[draw=s1,solid] table {plots/AIIterations-svcomp-Reach-32bit-Automizer_Default+AIv2_COMP_Simple_total.epf.plot};
\addplot[draw=s3,dashed] table {plots/AIIterations-svcomp-Reach-32bit-Automizer_Default+AIv2_INT.epf.plot};
\addplot[mark repeat={10},draw=s6,solid,mark=diamond] table {plots/AIIterations-svcomp-Reach-32bit-Automizer_Default+AIv2_INT_total.epf.plot};
\addplot[draw=s2,dotted] table {plots/AIIterations-svcomp-Reach-32bit-Automizer_Default+AIv2_OCT.epf.plot};
\addplot[mark repeat={10},draw=s4,solid,mark=star] table {plots/AIIterations-svcomp-Reach-32bit-Automizer_Default+AIv2_OCT_total.epf.plot};
\end{axis}
\end{tikzpicture}
	\phantomcaption
	\end{subfigure}
	\hfil%
	\begin{subfigure}[t]{0.49\textwidth}
	\centering
	\begin{tikzpicture}[scale=\plotscale]
\begin{axis}[%
xmin=0, ymin=0,%
xlabel={Samples},%
ylabel={Successful AbsInt Refinements},grid=major,%
legend style={at={(0.025,0.975)},anchor=north west,legend cell align=left}%
]%
\addlegendimage{empty legend}
\addlegendentry{Useful abs. interp. refinements}
\addplot[mark repeat={10},draw=s7,solid,mark=x] table {plots/AI-Refinements-svcomp-Reach-32bit-Automizer_Default+AIv2_COMP_Simple.epf.plot};
\addplot[draw=s1,solid] table {plots/AI-Refinements-svcomp-Reach-32bit-Automizer_Default+AIv2_COMP_Simple_total.epf.plot};
\addplot[draw=s3,dashed] table {plots/AI-Refinements-svcomp-Reach-32bit-Automizer_Default+AIv2_INT.epf.plot};
\addplot[mark repeat={10},draw=s6,solid,mark=diamond] table {plots/AI-Refinements-svcomp-Reach-32bit-Automizer_Default+AIv2_INT_total.epf.plot};
\addplot[draw=s2,dotted] table {plots/AI-Refinements-svcomp-Reach-32bit-Automizer_Default+AIv2_OCT.epf.plot};
\addplot[mark repeat={10},draw=s4,solid,mark=star] table {plots/AI-Refinements-svcomp-Reach-32bit-Automizer_Default+AIv2_OCT_total.epf.plot};
\end{axis}
\end{tikzpicture}
	\phantomcaption
	\end{subfigure}	
\caption{%
Statistics collected during the execution of the benchmarks.
All plots show the measured data on the y-axis and range over the samples on the
x-axis. 
The order of the samples is sorted by the measurement value for each plot.
This allows us to show trends but also prevents the comparison of single
samples.
The upper-left chart ``Runtime'' compares the total runtime of the different
settings. 
The upper-right chart ``Total refinements" compares the number of iterations,
the lower-left chart ``Abs. interp. refinements'' compares the number of
iterations were abstract interpretation was applied to path programs with loops,
and the lower-right chart ``Useful abs. interp. refinements'' shows the number
of refinements were abstract interpretation computed a proof for the
infeasibility of the path program.
 }
\label{fig:stats}
\end{figure}
In this section, we present the implementation and evaluation of our approach.

We implemented our algorithm in
\automizer\footnote{\url{https://ultimate.informatik.uni-freiburg.de/automizer}},
a state-of-the-art software model checker which is part of the \ultimate
framework\footnote{\url{https://ultimate.informatik.uni-freiburg.de}}.
\automizer uses the trace abstraction algorithm (see
Section~\ref{sec:traceabstraction}) and large block
encoding~\cite{fmcad/BeyerCGKS09}.

We also implemented an abstract interpretation engine in \ultimate.
Our engine supports octagons~\cite{DBLP:journals/lisp/Mine06} as relational
abstraction, intervals and congruences~\cite{DBLP:conf/tapsoft/Granger91} as
non-relational abstractions, and any combination thereof.
It also allows a union of different abstractions for each fixpoint, which can be
parameterized.

In our experiments, we used the same settings for the \automizer part of our
approach~\cite{DBLP:conf/tacas/HeizmannDGLMSP16} that were used when \automizer
participated in the software verification competition
\svcomp~2016~\cite{DBLP:conf/tacas/Beyer16}.
In particular, it uses Z3~\cite{MouraBjorner2008} to decide feasibility of a
sample trace in all our experiments.

For our evaluation we applied our version of \automizer to C programs taken from
the \svcomp~2016~\cite{DBLP:conf/tacas/Beyer16}
repository\footnote{\url{https://github.com/sosy-lab/sv-benchmarks/releases/tag/svcomp16}}.
We used the three sample sets ``Loops'', ``Simple'' and ``ControlFlow'',
which consist of a total of 237 benchmarks.
We removed all examples from the directory \texttt{ssh-simplified} (category
``ControlFlow''), because our abstract interpretation implementation contained a
bug that prevented it from running on those examples.
\reviewerins{Admitting that
your implementation is buggy does not convey a lot of confidence in the
results you obtained. You should probably fix the bug before submitting the
paper (and the results).}
This left us with 214 benchmarks.
\reviewerins{Participate in ``Loops'', ``Simple'', and ``ControlFlow'', but
\automizer didn't participate in SV-COMP'16. Why include this comparison?}
\reviewerins{If the implementation of the AI module contains a bug, how do you
know that the bug does not impact the other programs? Your results may be
biased because of the bug (the programs that you analyse faster could benefit
from the bug). As the main contribution of the paper is the experimental
section, this is a serious concern.}

\medskip

Each of the benchmarks contains one error location, which is either reachable or
unreachable. 
For 133 benchmarks, the location is unreachable, for 81 it is reachable.
The particular features of the programs of each category are as follows.
``Loops'' contains programs that contain multiple functions which each contain
multiple possibly nested loops that manipulate the variables of the program.
%
%
Note that programs in this category do not contain recursive function calls.
Category ``Simple'' contains programs obtained by simplifying real-world
examples. 
They contain multiple functions that are called from one single loop in the main
function, and use multiple different data structures such as enumerations,
structs, and unions.
The ``ControlFlow'' category consists of programs in which multiple control
variables are set within loops. 
The values of the control variables determines when a program enters an
erroneous state.
We compare the following seven different settings with each other:
\begin{itemize}[itemsep=0pt,topsep=0pt]
  \item \automizer without any modifications (\setAuto),
  \item our algorithm using an interval abstraction without data automaton
  enhancement (\setInt), or with construction of the enhanced data
  automaton (\setIntT),
  \item our algorithm using an octagon abstraction without data automaton
  enhancement (\setOct) or with construction of the enhanced data automaton 
 (\setOctT), and
  \item our algorithm using a combination of congruence and octagon abstraction
  without data automaton enhancement (\setComp) or with construction of the
  enhanced data automaton (\setCompT)
\end{itemize} 
All benchmarks were run on an Intel Core i5-3550 with 3.30GHz using a timeout of
90 seconds and a memory limit of 4GB for the tool itself and 2GB for the SMT
solver.

\begin{table}[t]
\centering
\begin{tabu} {lccc}
\toprule
\header{}         & \header{Success} & \header{Timeout} & \header{Error} \\
\cmidrule{1-4}
\header{\setAuto} & 97 (3)  & 110 &  7 \\
\header{\setOctT} & 97      & 106 & 11 \\
\header{\setOct}  & 100     & 106 &  8 \\
\header{\setCompT}& 101     & 100 & 13 \\
\header{\setComp} & 103     & 98  & 13 \\
\header{\setInt}  & 91      & 104 & 19 \\
\header{\setIntT} & 91      & 104 & 19 \\
\midrule                          
\header{Portfolio}& 108     &  89 &  6 \\
\bottomrule
&  & & \\
\end{tabu}
\caption{%
The evaluation results. 
The complete benchmark set contained 214 samples.  
Each cell in the column ``Success" contains the number of samples this
particular setting could solve.
The number in parenthesis shows how many samples were solved
\textit{exclusively} by this setting.
The column ``Timeout'' contains the number of times each setting run into the
90s timeout. 
The column ``Error'' contains the number of times each setting could not solve
a benchmark due to a crash of the tool. 
The row ``Portfolio'' shows how many benchmarks could be solved by any of the
settings, or on how many benchmarks all settings failed or errored, respectively.
}\label{tab:eval}
\end{table}
\reviewerins{The new algorithm solved 103 compared to the original's 97. This
is somewhat underwhelming. The paper itself mentions one factor which
contributes to this: ``Note that in nearly half the cases, [abstract
interpretation] was not necessary, because the benchmark could be solved
anlyzing only single traces.'' While the approach sounds promising, it seems
like more effort is needed to fully realize its potential.}

Table~\ref{tab:eval} shows the results of the evaluation. Out of the 214 input
programs, the default trace abstraction implementation was able to solve 97
programs.
The best abstract interpretation based CEGAR approach, \setComp, was able to
solve 103 programs.
The default configuration of \automizer (\setAuto) could solve three examples
exclusively.
These examples lead to timeouts in all the other settings. 

Figure~\ref{fig:stats} shows various statistics of all approaches compared to
each other.
The top left hand chart shows the runtime in $\log(s)$ for all individual
benchmark programs, ordered by time.
It shows that the \setComp{} setting was not only able to prove the most
programs, but also took the least time.
The fact that \setComp{} could solve the most problems was not unexpected: 
\setComp{} computes relational constraints of the form $\pm x \pm y \leq c$,
where $x$ and $y$ are variables and $c$ is a constant and combines them with
non-relational constraints of the form $x\!\! \mod c = 0$.
Both of these constraints are notoriously difficult to obtain for SMT solvers. 
Interestingly, the combination of octagons and congruence is even faster than
using octagons alone (\setOct{}). 
Although it is only a slight advantage, it shows that the additional
information is useful in some cases. 

The runtime chart also shows that the interval abstraction performed the worst
of all of our settings.
The reason for that is the missing precision of the interval domain compared to
the octagon domain. 
Therefore, more iterations to prove a program are needed and the timeout occurs
faster.

On the top right hand side in Figure~\ref{fig:stats} the number of refinements
in the CEGAR loop is shown.
This number indicates how often a new data automaton was constructed with either
the default trace abstraction algorithm or with the approach presented in this
paper.
Note that trace abstraction always needed to do more CEGAR iterations than the
\setOct, \setOctT, \setComp, and \setCompT{} settings.
Only the interval abstraction based settings were trailing the default trace
abstraction approach.
This fact shows that the choice of an relational abstraction (with combination
of the congruence abstraction) improves the convergence and the precision of the
overall approach.
Also note that about 30 programs could be proven in the first iteration of the
CEGAR loop.

The lower left hand chart of Figure~\ref{fig:stats} shows the number of
iterations in which a path program was constructed and analyzed with abstract
interpretation.
Note that in nearly half the cases, such a construction was not necessary,
because the benchmarks could be solved analyzing only single traces. 

The lower right hand chart shows the number of iterations in which abstract
interpretation could prove the infeasibility of the path program.
Compared to the total number of abstract interpretation refinements, in roughly
half the benchmarks this was the case.  
Interestingly, there is no sample for which the interval abstraction was useful
more than once, again outlining that the default variant of \automizer can infer
these invariants by itself.

The results in Table~\ref{tab:eval} and Figure~\ref{fig:stats} also show that
enhanced data automata do not perform better.
\reviewerins{This option turns out to not be an actual enhancement in any of
the cases; while there is something to be said for also presenting
unsuccessful ideas, I would have expected a more thorough investigation of why
this option did not produce any improvements.}
It seems that the checks required for adding additional edges take too much time
compared to their usefulness. 
\FloatBarrier

\section{Related Work}\label{sec:relwork}

\reviewerins{Another concern I have is the comparison with related work:
CPAChecker also uses AI. What is the difference/novelty in your approach? Can
you compare your results with CPAChecker on the SV-COMP benchmarks?}

In their work on Craig Interpretation~\cite{AlbarghouthiGurfinkelChechik2012,
DBLP:conf/cav/AlbarghouthiLGC12}, Albarghouthi et al. use a CEGAR-based approach
with abstract interpretation to refine infeasible program traces.
In contrast to our work, they use abstract interpretation to
compute an initial abstraction of the whole program.
Then, a trace to an error location is picked from the abstraction, instead of
the original program, and analyzed using a bounded model checker.
If the trace is infeasible, this results in a set of state assertions, which may
be too precise, i.e., non-inductive, to be used to refine the initial
abstraction.
Abstract interpretation is used again, this time to weaken the found state
assertions in an attempt to achieve inductivity before refinement of the last
abstraction is done and the next iteration begins.
Because the analysis is done on an abstraction dependent on the fixpoint
computed by abstract interpretation, many iterations are needed in the worst
case to identify infeasible program traces.
The fact that we are using abstract interpretation to compute fixpoints of path
programs which are a subset of the original program, instead of an abstraction,
allows us to circumvent the problem that an abstraction of the whole
program might be too weak to prove the program to be correct.
Additionally, we often eliminate the need to use expensive model checking
techniques to refine the abstraction iteratively.
Therefore, our generalization with abstract interpretation is more localized and
more precise than an abstraction obtained by analyzing the whole program.

Beyer et al. use path programs in a CEGAR approach to compute invariants of
locations in a control flow graph of a program~\cite{DBLP:conf/pldi/BeyerHMR07}.
The refinement of the abstraction is done by using a constrained-based invariant
synthesis algorithm which computes an invariant map, mapping predicates forming
invariants to locations of the path program.
Those invariants are excluding already visited parts from the original program.
This is done until a counterexample for the program's correctness has been
found or the program has been proven to be correct.
In contrast to our work, their approach uses an interpolant generator to
generate the invariant mapping, whereas we use both, an interpolant generator
and a fixpoint computation engine to obtain suitable state assertions.
In addition, their approach is only able to synthesize loop invariants by using
invariant templates which are parametric assertions over program variables,
present in each location of the program.
Although they propose to use other approaches to generate invariants, including
abstract interpretation, they do not present a combination of those methods.

\section{Conclusion}\label{sec:concl}
In this paper, we presented a CEGAR approach that benefits from the precision of
trace abstraction and the scalability of abstract interpretation.
We use an automata theoretical approach to pick traces from a program automaton
which are checked for infeasibility.
If the trace is infeasible, we construct a path program and compute an
abstraction of the path program by using abstract interpretation.
With the help of this abstraction, we are guaranteed to obtain state assertions,
in particular loop invariants, which help us to exclude a generalization of the
found infeasible trace from the program.
Because abstract interpretation may yield an abstraction which is not precise
enough to synthesize usable loop invariants, we use the default precise trace
abstraction approach as a fallback.

Our experiments show that by using abstract interpretation to generate loop
invariants of path programs, we are to not only able to prove a larger set of
benchmark programs, but also need less CEGAR iterations to do so, leading to a
more efficient approach to proving correctness of programs.

\FloatBarrier

\bibliographystyle{abbrv}
\bibliography{main}

\setcounter{tocdepth}{1}

\end{document}